# Fast Activation of Graphene with Corrugated Surface and its Role in Improved Aqueous Electrochemical Capacitors

*Longsheng Zhong,[1,#] Chang Wu,[2,3,#] Xiaojing Zhu,[4] Shulai Lei[5,*] Guijie Liang,[5] Sepidar Sayyar,[6,7] Biao Gao,[1,*] and Liangxu Lin[1,3,7,*]*

[1]Institute of Advanced Materials and Nanotechnology, The State Key Laboratory of Refractories and Metallurgy, Wuhan University of Science and Technology, Wuhan 430081, China
[2]Australia Institute for Innovative Materials, Innovation Campus, University of Wollongong, Squires Way, North Wollongong 2519, Australia
[3]The Straits Laboratory of Flexible Electronics (SLoFE), Fujian Normal University, Fuzhou 350117, China
[4]School of Mechanical and Electronic Engineering, China Jiliang University, Hangzhou 310018, China
[5]Hubei Key Laboratory of Low Dimensional Optoelectronic Materials and Devices, Hubei University of Arts and Science, Xiangyang 441053, China
[6]Australian National Fabrication Facility – Materials Node, Innovation Campus, University of Wollongong, Wollongong, NSW 2519, Australia
[7]ARC Centre of Excellence for Electromaterials Science, Intelligent Polymer Research Institute, Australia Institute for Innovative Materials, Innovation Campus, University of Wollongong, Squires Way, North Wollongong 2519, Australia

[#]Contribute equally to this work
*Correspondence: l.lin@wust.edu.cn (L. L.); sllei@hbuas.edu.cn (S. L.); gaobiao@wust.edu.cn (B. G.)

**ABSTRACT**

In graphene based materials, the energy storage capacity is usually improved by rich porous structures with extremely high surface area. By utilizing surface corrugations, this work shows an alternative strategy to activate graphene materials for large capacitance. We demonstrate how to simply fabricate such activated graphene and how these surface structures helped to realize considerable specific capacitance (*e.g.,* electrode capacitance of ~340 F $g^{-1}$ at 5 mV $s^{-1}$ and device capacitance of ~ 343 F $g^{-1}$ at 1.7 A $g^{-1}$) and power performance (*e.g.,* power density of 50 and 2500 W $kg^{-1}$ at the energy density of ~10.7 and 1.53 Wh $kg^{-1}$, respectively) in aqueous system, which are comparable to and even better than those of highly activated graphene materials with ultra-high surface area. This work demonstrates a new path to enhance the capacity of carbon-based materials, which could be developed and combined with other systems for various improved energy storage applications.

## 1. Introduction

As one of the key clean energy technologies, the electrochemical capacitor (ECs, supercapacitor, mainly based on carbon materials) performs fast and stable energy storage kinetics, but has the critical shortage of low energy density (capacitance) [1-3]. Improvement on the capacitance has been realized by adding pseudo-capacitive materials, which usually sacrifices the power density and life time [4-8]. In this regard, developments of all carbon based ECs are attractive if their capacity can be suitably enhanced [9-17].

In general, the capacitance of ECs is mainly contributed by electrochemical double-layer (EDL) ion adsorption at Helmholtz layer, which is decided by the surface area and electrolyte dielectric constant [1,3]. The Brunauer–Emmett–Teller (BET) surface area and rich porous structure were then considered highly critical in improving the capacitance of electrode materials. For example, a very high capacitance of 150 F $g^{-1}$ (at 2.7 V and current density of 0.8 A $g^{-1}$) has been demonstrated on the KOH activated graphene with extremely high BET surface area of ~ 3100 $m^2$ $g^{-1}$ [9]. Nevertheless, recent progress suggests that the ultra-high BET surface area may not always obligatory for large capacitance. Compared to the 2D graphene plane, edge-like structures are also highly crucial in improving the capacitance by giving a much higher solvent accessible surface area (SASA, *e.g.,* SASA of 1.4 at the zigzag edge *versus* 1 of the plane, the SASA reflects the corrugation structures and associated with the density of ions in EDL) and thinner Helmholtz layer [3,13,18,19]. Such improvement is reflected by the increased binding energy to electrolyte ions (*note: the ion adsorption distance is reduced, the Helmholtz layer is compressed with higher concentration of adsorbed ions*), and arose from the enhanced density of state (DOS) around the Fermi level [3,18-21]. Along with the enhanced DOS at the Fermi level, the quantum capacitance at local region should be also elevated [3]. The topological defects (*e.g.,* corrugations/crumples) may have similar effect to that of edge-like structures to improve DOS at the Fermi level and enhance binding abilities [21-26], although interpreting these effects on the capacitance from theoretical standpoint is still challenging.

This work discusses the above hypothesis and demonstrates how graphene materials can be quickly activated with corrugated surfaces. Although the activated graphene has a low BET surface area of ~340 $m^2$ $g^{-1}$, it still exhibited comparable energy storage performance to that of highly activated graphene with ultra-high BET surface area (*e.g.,* ~3100 $m^2$ $g^{-1}$). Combing with theoretical calculations, we suggest why and how the graphene can be activated by means of corrugations/crumples, which offers a simple way to improve the energy storage capacitance of carbon based materials.

## 2. Experimental
### 2.1 Preparations
The graphene was activated from the graphene oxide (GO) film with a fast thermal expansion technique (**Figure 1**). GO suspension (800 mL from 4 g graphite flakes, Nacalai Tesque, product no. of 17346-25) was prepared with a modified Hummers method [27-29]. 12 mL GO suspension was poured onto a nitric acid fiber microporous filtration membrane in the glass culture dish (diameter of 7 cm, **Figure 1a**), and dried at 95 °C in air dry oven for 1 h to form the GO film. Ethanol was used to sock and separate the film from filtration membrane. The GO film was air-dried overnight (**Figure 1c**) and cut to small pieces (< 0.5×0.5 cm, no strict requirements on the shape, **Figure 1c** inset). GO pieces were then dropped into a quartz beaker (325-425 °C, heated by a hot plate), and quickly expanded upon contact to the beaker bottom (**Figure 1d**). Instantaneous burning (very short time) of the GO film was occasionally found in the beaker (**Figure 1d**). The expanded sample was maintained in the beaker for ~5 minutes. In case that some GO pieces were not fully expanded, the collected powder sample was simply ultrasonicated with ethanol solvent to remove solid sediments (the un-expanded graphene was on the bottom of glass vessel). The activated graphene was collected (centrifuged and dried) from the suspension, which have many nanoscale corrugations on the surface (will be discussed). For comparison, two control samples were also prepared to exclude surface corrugations. During our fabrications, we found that the efficient expansion highly relied on the thin film thickness (will be further discussed). Therefore, 50 mL GO suspension was ultrasonicated initially to fully exfoliate GO layers. 25 mL GO suspension was poured onto a nitric acid fiber microporous filtration membrane in the glass culture dish (diameter of 5 cm) to form a thicker film. With identical procedure, the GO film was transferred and cut to small pieces. These GO pieces (< 0.5×0.5 cm) were transferred into the quartz beaker which was gradually heated to 150 $^oC$ (maintained for 0.5 h), 200 $^oC$ (maintained for 0.5 h), 300 $^oC$ (maintained for 0.5 h) and then 375 $^oC$ (maintained for 5 minutes). The collected sample is denoted as **rGO-control-1** (r: reduced). It should be noted that no quick expansion of the GO film was found during this preparation. Besides, another 25 mL GO suspension was mixed with 1 mL hydrazine anhydrous (98%) and stirred for 2 h, which was then dried in the oven at 375 $^oC$ for 2 h. The collected sample is denoted as **rGO-control-2**.

## 2.2 Characterizations

Scanning electron microscopy (SEM) was performed with a Nova 600 NanoSEM scanning electron microscope. Transmission electron microscopy (TEM) and aberration corrected scanning TEM (AC-STEM) images were characterized with JEOL JEM-2010 (200 kV) and JEOL ARM 200F (80 kV), respectively. X-ray photoelectron spectroscopy (XPS) was performed on a Thermo Fisher Nexsa System. An ion beam etch was applied before the characterization, while the C1s was calibrated as 284.6 eV as a standard. Raman spectra were recorded by the Horiba LabRAM HR Evolution at 633 nm. During Raman characterizations, the beam power was controlled suitably to exclude the effect from thermal reduction of graphene based materials. $N_2$ and $CO_2$ adsorption/desorption isotherms were recorded with BSD-PS2 *via* Beishide Instrument Technology (Beijing) Co., Ltd. Contact angle was recorded with a micro-camera by dropping water on pressed graphene materials.

## 2.3 Electrochemical measurements

In three-electrode system, electrochemical measurements were conducted in 6M KOH with graphite rod and Hg/HgO (filled with 1M KOH) as counter and reference electrodes, respectively. The working electrode was fabricated by casting graphene powder (~1.5 mg) between nickel foams (1×1 cm). In two-electrode system, 6M KOH and 1M Li-PF6 were used as aqueous and organic electrolytes, respectively. In aqueous ECs, the electrode was made by sandwiching graphene powder (~0.8 mg for each electrode, *note*: the graphene power was too light and highly loosed, therefore only 0.8 mg graphene was used in the electrode fabrication) between nickel foam plates (diameter of 12 mm), and pressing (10 MPa) additional layers of nickel foam at both sides to prevent the leakage of graphene powder (**Figure 1b**). In organic ECs, graphene (~0.8 mg for each electrode), super P and polyvinylidene fluoride (PVDF) were mixed with suitable N-methyl-2-pyrrolidone (NMP) with a mass ratio of 8:1:1 to form a paste which was then pasted on aluminum foil, dried and punched into round thin film electrodes (diameter of 12 mm). The cell was assembled in CR2016 coin cell with the separator of porous polypropylene/PP film. Cyclic voltammetry (CV) and galvanostatic charge/discharge (GCD) curves and electrochemical impedance spectroscopy (EIS) were recorded by a CHI 660E electrochemical workstation. Energy storage performance of the cell was evaluated with a Neware battery cycler (CT-4008-5V10mA-164, Shenzhen, China). Capacitance was calculated with both CV and GCD curves (details in **Supporting Information**). In any CV curves, the calculation of capacitive and diffusion controlled factions was based on the well documented method (details in **Supporting Information**) [20,30,31].

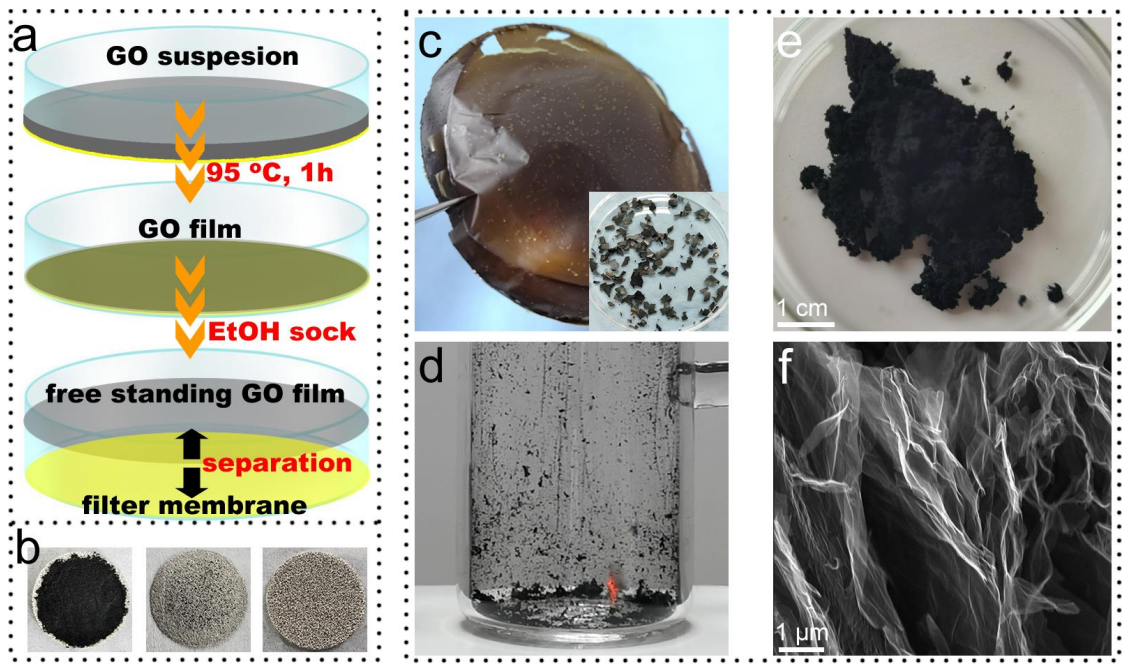

**Figure 1.** **(a)** The formation of GO film. **(b)** Nickel foam plates with graphene materials. From left to right are graphene materials on one nickel foam plate, sandwiched by two nickel foam plates and packed with additional nickel foam plates. **(c)** One dried GO film (diameter: 7 cm) and small GO pieces (inset). **(d)** The thermal expansion of GO pieces in quartz beaker. **(e)** The collected a-G-375 powder. (f) SEM image of a-G-375.

## 3. Results and discussion
### 3.1 Fabrication and characterization

In this fabrication, the thickness of GO film was highly crucial towards efficient activation of graphene sheets. As shown in **Figure 1a**, different GO films were prepared from various amount of GO suspension. After many attempts, the use of 12 mL GO was optimized, giving the best thermal expansion (**Figure 1d**). With the thicker film (*e.g.*, film formed from 18 mL GO suspension), the thermal expansion was suppressed which can be visually observed with some GO film just shrunk and twisted rather expanded. Preparation with fewer GO was also attempted, giving more difficulties to form homogeneous thickness of the film, which was difficult to be transferred and therefore was excluded in this work. As we described above, the GO film was cut to small pieces (< 0.5×0.5 cm, **Figure 1c** inset), to allow them being able to efficiently contact with the bottom of quartz beaker, and the self-propagation thermal expansion (the entire GO piece can be fully expanded upon the contact of any regions of the GO piece with beaker bottom). This is a simple expansion process, even when the later GO pieces were drop onto the surface of previously expanded rGO powder. There was no strict requirement on the shape and drop method of small GO pieces, *i.e.* drop one by one or drop tens of pieces in one time with tweezers (with multiple fabrications; if possible, the dropped GO pieces were simply stirred with scraper in the latter case to allow full expansion reactions). This conclusion was supported by our various CV scans and long-term development of the fabrication technique for comparison, *i.e.* the visible thermal expansions, CV shapes and capacitances were all close to each other except few wrong preparations. Controlling the temperature of hot plate yielded different activated graphene materials which are denoted as **a-G-*x*** (*x* means the temperature/°C). The thermal expansion was difficult to happen with the temperature < 325 °C, whilst it was uncontrollable upon the temperature > 425 °C due to the fast burning of graphene. We have attempted to use $N_2$/Ar to avoid burning, but was not really successful. The underlying reason of the burning of GO pieces was likely due to the fast thermal reaction (the thermal transport in such thin and small GO pieces are highly efficient) between carbon atoms and O-containing groups at high temperatures. Similar phenomenon was also found by Ruoff *et al.* in their thermogravimetric analysis of GO powder under $N_2$ atmosphere [32]. Therefore, the temperature for thermal expansion here was controlled between 325-425 °C.

The reported activations of graphene materials are varied from each other, which mainly involved the generation of rich porous structure (*e.g.*, activated by the reaction with KOH, the use of template and multiple thermal annealing) [9, 33-37]. The thermal reduction of GO was also well documented, to reduce GO and/or generate 3D porous structure [34,38,39]. Unlike traditional thermal annealing technique (*e.g.*, the annealing reaction in a furnace [34]), our developed fast fabrication technique has the potential for large-scale uses, with just hot plate and heating container. Most importantly, graphene sheets can be simply engineered with corrugated surface (will be further discussed) with this technique, which cannot be realized by other fabrications described above [9,33-39]. Unlike porous structures with high BET surface area, we show below a different landscape on the activation of graphene materials by using such corrugated surface to improve energy storage performances in aqueous ECs.

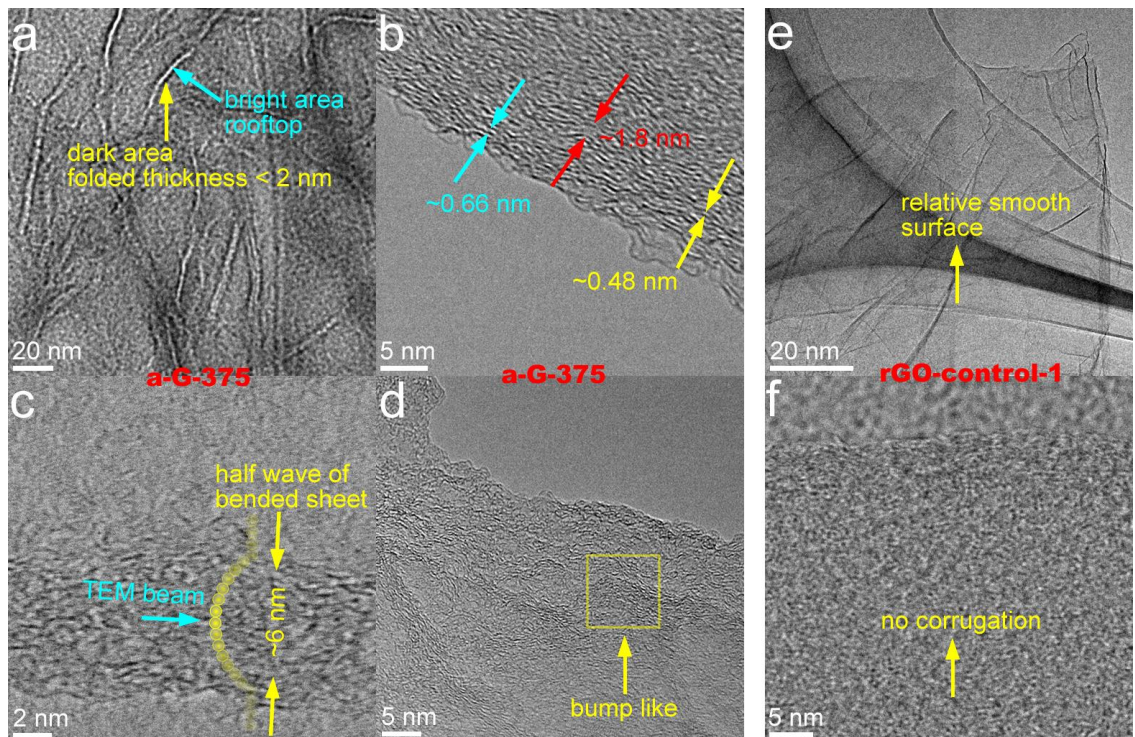

**Figure 2. (a-d)** Bright field TEM images of a-G-375. **(e-f)** Bright field TEM images of rGO-control-1. No clear surface corrugation was found on the rGO sheet, no matter the tilting of TEM holder was applied or not applied. *Note: the contrast difference in some TEM images (e.g., figure e) was induced by our CCD camera in JEOL JEM-2010, which has also been clarified in our previous work* [40].

Next, we focus on a-G-375, as an example, to show the structural information of activated graphene materials. The a-G-375 was highly loosened and light (~ 40 mg in **Figure 1e**). It was generally well reduced during the thermal expansion, giving a small portion of O-containing groups (carbon mole ratio of ~ 84.34%, **Table S1**, Raman and XPS characterizations in **Figure S1a-c**, the shape of the C1s XPS spectrum is similar to that of highly reduced graphene oxide material [28]). The activated graphene material has expanded structure, with graphene sheets highly bended and folded (**Figure 1f** and **Figure S2a**). During our TEM observations, the protruding regions of wrinkles were in focus to allow the imaging of layered information, while the other basal plane regions were defocused (without fine structure). **Figure 2a** shows the TEM image of such wrinkled positions (in some regions, not universal) with thickness < 2 nm (distance between two dark area was around 3.69 nm). The thickness of one layer sheet should be < 0.835 nm supposing a minimal distance of 0.33 nm (the interlayer distance of graphite) between two folded sheets, which corresponds well with the thickness of 1-3 layer graphene (*note*: 4 layer graphene has the thickness > 0.99 nm). These graphene sheets were highly corrugated (**Figure 2b, Figure S2b)**, giving widespread crumple structures with full wave of 0.4-1.8 nm (*note*: the structure is similar to waved graphene with periodic waved corrugations) [41]. Such surface corrugation were widespread on rGO sheets and was further revealed by the top view TEM image on the bended area with highly disordered lattices (incident TEM beam from the top of the bended position, **Figure 2c**; again, the rooftop of the bended area was in focus). More close examinations suggested that the activated graphene sheet has also many bump-like structures (formed from various corrugations), which are presented as bright dots in the dark-field scanning TEM image (**Figure S2c**, the erupted structures are in focus; energy-dispersive X-ray (EDX) spectrum of these structures only showed C and O elements, **Figure S2c** inset; the bump-like structure was integrated with and from the rGO sheet, **Figure S2d**). These bump-like structures (formed from various corrugations) are more evident in **Figure 2d** where the sheet was angled with the perpendicular direction of TEM beam. It is known that various O-containing groups and water molecules existed on the surface and in the stacked interlayers of GO sheets [27,28,32]. In the fast thermal expansion of thin GO films, the majority O-containing groups were removed while the water molecules were vaporized rapidly (generated structural asymmetry), constantly generating structural distortion and strain (surface corrugations/crumples) on the rGO surface. Such corrugation is common in graphene materials because of the anisotropic strain relaxation [42-44]. By contrast, no thermal expansion was found on the thicker GO film and multiple thermal reactions (**rGO-control-1**). TEM examinations of this control sample only showed thin and smooth surface of rGO sheets (**Figure 2e,f**), no matter the tilting of TEM holder was applied or not applied to adjust the focus on the fine structure.

More surface structures of a-G-375 were characterized by nitrogen adsorption/desorption at 77.3 K. The isotherm (**Figure 3a**) illustrates type H3 hysteresis (see IUPAC classification) [45], giving only gradually increased adsorption at medium and high pressure regions. It means that the micropore in a-G-375 was negligible, while the $N_2$ adsorption was mainly contributed by meso- to macrospores according to the general analysis. The a-G-375 has a big pore volume of ~2.46 mL $g^{-1}$, which is even higher than those (2.14 mL $g^{-1}$) of KOH activated graphene materials with ultra-high BET surface area of 3100 $m^2$ $g^{-1}$ [9]. Nevertheless, the BET surface area of the a-G-375 was calculated (from the linear relative pressure range from 0.04 to 0.32) to be only ~340 $m^2$ $g^{-1}$. The a-G-375 was also characterized with $CO_2$ adsorption at 273.15 K (**Figure 3a** inset). The cumulative pore volume and pore size derived from both $N_2$ and $CO_2$ isotherms are shown in **Figure 3b**, which again suggested that the adsorption was mainly contributed by meso- to macrospores. Since the Barrett-Joyner-Halenda (BJH) method is not applicable to macropores, the most probable pore size of 4.1 nm (with model of cylindrical pore geometry, average pore size of 14.4 nm) should be considered as an equivalent pore size only. The pore size of micropore analyzed from the $CO_2$ isotherm is around 0.75 nm (Horvath-Kawazoe method), although it only contributed few pore volumes (0.14 mL g$^{-1}$). With same characterizations, the annealed a-G-375-100 (at 1000 $^o$C for 2 h, denoted as a-G-375-$x$ where the $x$ represents annealing temperatures) was also analyzed, which has nearly the same surface structure to that of a-G-375 (**Figure 3a,b**). During TEM observations, no evident porous structures were found on a-G-375. The above gas adsorption (*e.g.*, the pore volume is larger than that of activated graphene materials with rich porous structure and high BET surface area, 3100 m$^2$ g$^{-1}$) may be also contributed by the improved gas adsorption on the surface which is associated with the enhanced DOS at the Fermi level (will be further discussed combing with theoretical calculations) and have been well discussed in literatures (usually arose from the point defect, structural asymmetry, lattice dislocation/relaxation and surface strain) [21,26,46,47].

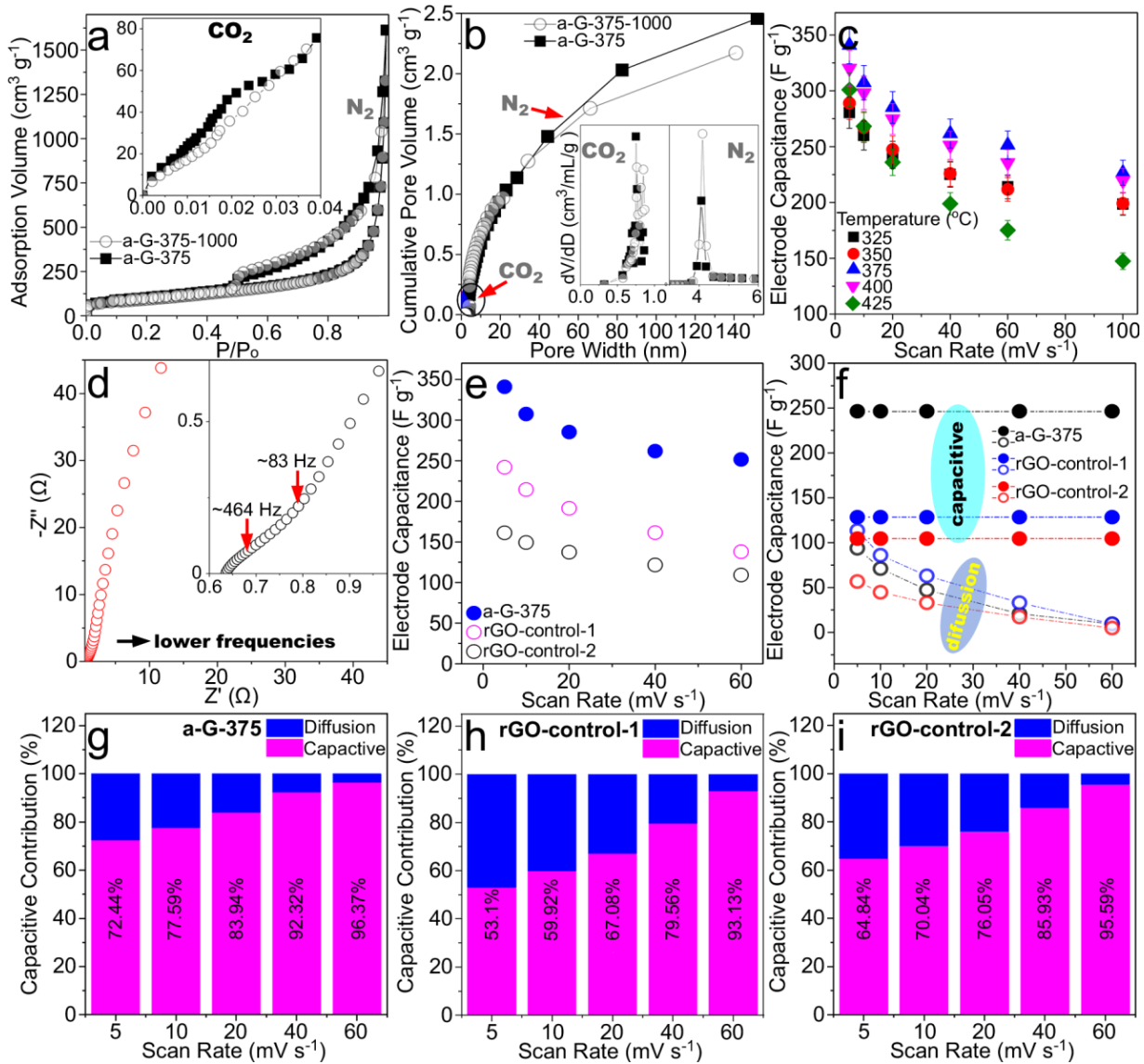

**Figure 3.** (**a**) N$_2$ and CO$_2$ adsorption/desorption isothermal plots of a-G-375 and a-G-375-1000. (**b**) Cumulative pore volume and pore size distribution of a-G-375 and a-G-375-1000. (**c**) Electrode capacitance of a-G-375.

The error of the capacitance was treated as 5%. **(d)** The Nyquist plot shows the imaginary part versus the real part of impedance of a-G-375. Inset is the plot at high frequency ranges. **(e)** Comparison of the electrode capacitance of a-G-375, rGO-control-1 and rGO-control-2 at different scan rates (see Figure S6 for CV curves of control samples). **(f)** Comparison of the capacitive and diffusion controlled contribution to the entire electrode capacitance of a-G-375, rGO-control-1 and rGO-control-2. **(g-i)** The capacitive contribution at different scan rates of a-G-375 (g), rGO-control-1 (h) and rGO-control-2 (i). *Note: above electrochemical measurements were performed in 6M KOH with three-electrode technique. For the calculation of capacitive fraction, scan rates of 5, 10, 20, 40 and 60 mV s$^{-1}$ were investigated. CV curves with higher scan rate were excluded in this calculation.*

## 3.2 Electrochemical performance

Although the BET surface area of a-G-375 was not high enough, it still exhibited large capacitance in 6 M KOH electrolyte. Compared to other a-G-*x*, the a-G-375 has the highest electrode capacitance of ~340 F g$^{-1}$ at 5 mV s$^{-1}$ (roughly equals to 1.7 A g$^{-1}$, **Figure 3c**, CV curves in **Figure S3a-e**), which is very high in graphene materials. In spite of the low mole ratio of O-containing groups (**Table S1,** carbon ratio of ~ 84.34%, close to that of chemical reduced graphene oxide [28] and other highly reduced graphene materials, see **Table S2**), the pseudo-capacitance responses in the CV curve (6 M KOH, from 0 to -1.0 V versus RHE, **Figure 3f**, **Figure S3a**, three-electrode) of a-G-375 are still evident. These electrochemical responses were likely originated from surface redox reactions of C-OH↔C=O, C=O↔CO$^-$ and COOH↔COO [48,49]. It was also reflected by the electrochemical impedance spectroscopy (EIS) in **Figure 3d**, as an ideal capacitive process should have a nearly vertical plot at lower frequencies (zero resistance) [50]. In this EIS, a transition between the RC semicircle and electrolyte migration was observed at the frequency of ~464 Hz. The diffusion of electrolyte changed at ~83 Hz, with contributions from pseudo-capacitance (**Figure 3d**). It is reasonable to argue that the high capacitance of a-G-375 may be mainly contributed by the pseudo-capacitance from O-containing groups. Nevertheless, the following control experiments and comparisons suggest that this is not the case, whilst the corrugation structures have to be further considered as a main contributor.

To investigate the capacitive and diffusion controlled (pseudo-capacitive) contributors of a-G-375, two control samples were also fabricated (rGO-control-1 and rGO-control-2) for comparison. Graphene sheets in these control samples were all fully exfoliated as that showed in TEM images (**Figure 2e** and **Figure S4a**). With careful TEM observations, surface corrugations were excluded in these control samples (**Figure 2e-f**, **Figure S4**). XPS analyses suggested that rGO-control-1 and rGO-control-2 have C ratios of 74.3 and 79.7%, respectively (*versus* 84.34% of a-G-375, **Figure S1** and **Table S1**). Although a-G-375 has fewer O-containing groups on the surface, it still exhibited much higher electrode capacitance (~340 F g$^{-1}$ at 5 mV s$^{-1}$) over that of control samples (**Figure 3e**). Without surface corrugations, rGO-control-1 and rGO-control-2 have only electrode capacitances of 241.8 and 161.1 F g$^{-1}$ (scan rate of 5 mV s$^{-1}$), respectively. With the established method (details in **Supporting Information**) [20,30,31], the capacitive and diffusion controlled contributors of CV curves can be plotted (*e.g.,* **Figure S5, Figure S6a,b**), and their contributions to the entire capacitance were calculated. Results of this comparison were summarized in **Figure 3f**, and detailed in **Figure 3g-i** (also see **Figure S6** for other a-G-*x* and a-G-375-1000 samples). The fraction of capacitive contributor in all a-G-*x* (*x*=325, 350, 400, 425) are close to that of a-G-375 (**Figure 3g**, **Figure S6**). rGO-control-1 and rGO-control-2 have the capacitive controlled capacitance of 128.4 and 56.6 F g$^{-1}$ (at the scan rate of 5 mV s$^{-1}$; pseudo-capacitance of ~113.4 and 56.6 F g$^{-1}$), respectively (**Figure 3f**). With the presence of surface corrugations, the capacitive controlled capacitance of a-G-375 was significantly improved to

~246.3 F g$^{-1}$ (72.46% of 340 F g$^{-1}$, scan rate of 5 mV s$^{-1}$; diffusion controlled contribution of ~93.7 F g$^{-1}$). This value is already higher than that of highly activated graphene materials with rich porous structure and super higher BET surface area (~3100 m$^2$ g$^{-1}$, ~150 F g$^{-1}$, two-electrode system) [9], even when a larger potential window (~2.7 V) was applied in the latter case (two-electrode measurement of the cell fabricated from a-G-375 will be further discussed below in **Figure 4**, which has similar device capacitance of 342.7 F g$^{-1}$). Therefore, in a-G-375, the most capacitance (*e.g.,* 72.46% at the scan rate of 5 mV s$^{-1}$) was contributed by the capacitive energy storage, while the presence of corrugations has greatly improved the capacitance.

The few O-containing groups (see XPS in **Figure S1b** and **Table S1**) made a-G-375 highly hydrophilic (quickly wetted by water, **Figure 4a** inset). By contrast, the annealed a-G-375-1000 (annealed for 2 h) has fewer O-containing groups (**Figure S1b**, carbon ratio of ∼ 90.4%, -OH were fully removed) and was highly hydrophobic, giving large water contact angle (**Figure 4a inset**) and the floating of some graphene particles on the surface of water drop (more details in **Figure S1f**). These O-containing groups are unavoidable in carbon materials, especially in those materials with rich defect structures (see **Table S2**). Correspondingly, the capacitance of annealed a-G-375 gradually reduced along with the annealing temperature (**Figure 4a**, calculated from the CV curves at scan rate of 5 mV s$^{-1}$ in **Figure S3f**). Compared to a-G-375, the annealed samples (all annealed for 2 h at different temperatures) exhibited better rate performance (**Figure S7**). The hydrophilicity of materials should have significantly affected their capacitance and rate performance as the carbon ratios in both a-G-375 and a-G-375-1000 (mole ratio of ∼ 84.34% and 90.4%, **Table S1**) were actually close to each other. To the best of our knowledge, the surface energy of activated graphene materials can be changed by both surface functional groups (*e.g.,* -OH) and corrugations, while the defect improves surface energy accordingly [27,51]. According to the surface energy matching theory [27,52], the corrugated graphene with increased surface energy should be more suited for aqueous electrolytes comparing to pure graphene materials.

Therefore, we further assembled the aqueous symmetric EC with a-G-375 (electrolyte of 6M KOH) and estimated the energy storage performance. As shown in **Figure 4b**, the cell has the device capacitance of 342.7, 297.7, 247.1, 196.3, 161, 121.8 and 80.9 F g$^{-1}$ at scan rate of 5 (roughly equals to 1.7 A g$^{-1}$), 10, 20, 40, 60, 100 and 200 mV s$^{-1}$, respectively. These values are much larger than that reported on various graphene materials, including the exfoliated graphene, graphene materials with rich pseudo capacitance and highly activated graphene materials with ultra-high BET surface area and similar O contents (**Figure 4c**, details in **Table S2**, full list of references in **Supporting Information**). The capacitances calculated from the GCD curves are slightly lower than those from CV curves due to possible underestimation of pseudo-capacitance in GCD curves (**Figure 4d**). With GCD curves, the power density of device was calculated as 50 and 2500 W kg$^{-1}$ at the energy density of ~10.7 and 1.53 Wh kg$^{-1}$, respectively. These values are also greater than those from other similar graphene materials, and comparable to highly activated graphene materials (*note:* with similar O contents from the C1s XPS spectrum) with extremely high BET surface area (*e.g.,* ~2000 m$^2$ g$^{-1}$, **Figure 4e,** details in **Table S2**). With the chronopotentiometry (CP) method, the cyclic performance of the symmetric EC was also evaluated. Show as **Figure 4f** is the result of this measurement, which illustrated a ~5% and 7.6% loss of the energy density at cycle numbers of 5000 and 10000, respectively. The electrochemical stability may be not high enough, but can be considered highly promising in aqueous system for long-term run. Further analysis of the system suggested that the decay of energy density was mainly induced by the anode, where the activated carbon structure maybe too active at high potentials (details in **Figure S8**, **Supporting Information**).

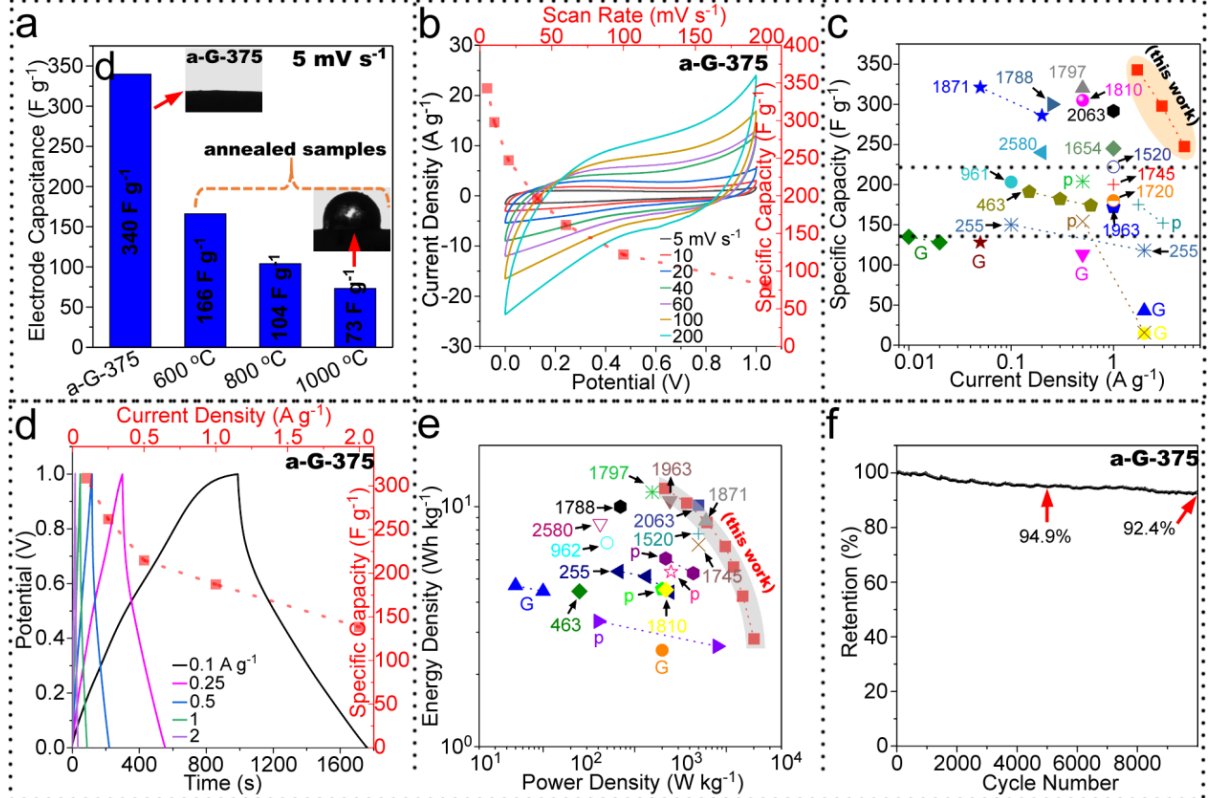

**Figure 4. (a)** Electrode capacitance of a-G-375 samples annealed at different temperatures (three-electrode). Insets show the water contact angle of a-D-G and annealed a-G-375-1000. **(b)** CV curves and associated specific capacitance plot of the cell. **(c)** Comparison of the device capacitance with other reported graphene based materials. **(d)** GCD curves and specific capacitance plot. **(e)** Ragone plot and the comparison of the power/energy density to other reported graphene materials. In (c) and (e), the G and p represent relatively pure graphene and graphene materials with rich pseudo-capacitance, respectively. The digital number illustrates the BET surface area (m$^2$ g$^{-1}$). **(f)** Cyclic test of the cell at the current density of 0.5 A g$^{-1}$ with CP method. In (b)-(f), the cell device was assembled from a-G-375 with 6M KOH electrolyte.

With suitable annealing treatment, the activated graphene materials can be tuned to relatively suit organic electrolytes, *i.e.* a capacitance of 137.4 F g$^{-1}$ (annealed at 1000 $^{o}$C for 2 h) can be reached at the scan rate of 5 mV s$^{-1}$ (roughly equals to charge/discharge rate of 0.25 A g$^{-1}$) in 1M LiPF$_6$ (potential window of 2.7 V), which is close to the highly activated graphene materials (150 F g$^{-1}$) with ultra-high BET surface area (~3100 m$^2$ g$^{-1}$) at similar potential window [9]. Nevertheless, comparison between the energy storage performances in both aqueous and organic electrolytes (see **Figure S9-10** and **Supporting Information** for more details) suggested that current activated graphene materials are more suited for aqueous media. As we illustrated above, the corrugation structure, in principle, improves the surface energy of graphene materials, making them more compatible to aqueous system (with higher surface tension over that of organic electrolyte; such "surface energy matching theory" has been well discussed previously [27,52]). Future directions in this regard may involve the modification of activated graphene materials to adjust their wettability. Besides, to further promote the activated graphene in energy storage, modifications to improve the chemical stability of anode materials should also be considered (*note*: in electrochemical oxygen evolution reactions, the catalyst electrodes were usually unstable at high working potentials; redox potential of highly activated carbon materials may be reduced, which is reasonable owing to the improved Fermi level DOS and chemical activity [21]).

## 3.2 Theoretical understanding

As we know, the energy storage capacitance at the EDL ($C_{EDL}$) can be calculated as [3]:

$$C_{EDL} = \frac{\varepsilon \cdot A}{d} \quad (\textbf{Eq. 1})$$

where $C$, $\varepsilon$, $A$, and $d$ are the capacitance, electrolyte dielectric constant, solvent accessible surface area (SASA), and the separation between electrolyte ions and surface of electrode materials, respectively. The total capacitance of electrode ($C_{tot}$) is determined by the quantum capacitance ($C_q$) and liquid electrolyte as follows [19,53]:

$$C_{tot}^{-1} = C_q^{-1} + C_{EDL}^{-1} + C_{Dielec}^{-1} \quad (\textbf{Eq. 2})$$

where the $C_{Dielec}$ is the dielectric screening of electrode. Furthermore, the $C_q$ is the intrinsic property of electrode material related to the electronic structure *via* the following equation [54]:

$$C_q(V) = \frac{1}{S \cdot V} \int_0^V eD(\varepsilon_F - eV') dV' \quad (\textbf{Eq. 3})$$

where $S$, $V$, $D$, $\varepsilon_F$, and $e$ are the surface area, applied voltage, DOS, Fermi level, and electron charge, respectively. Therefore, the SASA, distance between electrolyte ions and surface of electrode materials, and DOS at the Fermi level are considered highly crucial in improving the electrode capacitance in electrolytes.

To understand how the corrugation improves energy storage capacity through the change of above three factors, density functional theory (DFT) calculations were performed by using the Vienna *ab initio* Simulation Package (VASP). Cutoff energy of plane-wave basis was set to 500 eV to treat the electron-ion interactions in projected augmented wave pseudopotential method. The Brillouin Zone was sampled with a 1×7×1 k-point grid and converging tolerance of 0.01 eV/Å for residual force during geometry relaxation, while it was sampled with a 1×9×1 k-point grid and converging tolerance of $10^{-6}$ eV for energy during electron structure analysis. In order to prevent the interaction of periodic images, a vacuum space in the z-direction was set to 20 Å. The van der Waals (vdW) interactions were described by the semi-empirical DFT-D2 method. The corrugation ratio $R$ was defined as $R/100 = (L_0-L)/L_0$, in which $L_0$ is the lattice constant of plane graphene and $L$ for corrugated graphene, respectively. The wave length in crumples was measured as ~0.4-1.8 nm (**Figure 2b** and **Figure S2b**, the majority), which increased to around < 3.69 nm in bended positions (**Figure 2a**). Therefore, the $L_0$ was set as 36.9 Å supposing a non-corrugation region. Five values of the L (L = 27, 29, 31, 33 and 35 Å) were chose to cover different corrugation extents ($R$ ratio up to 26.8%, **Figure 5a**), giving the trend of surface binding ability associated with the corrugation extents. Higher R ratios (*e.g.,* fully bended) were excluded as which would tend to breaking the structure which is not the purpose of our calculations.

**Figure 5a** shows different optimized graphene with $R$ = 0, 5.1, 10.6, 15.9, 21.4, 26.8. In these structures, the nearest six C ring to adsorb ions (*e.g.,* Li$^+$ or K$^+$), including two different sits of carbon atoms (C1 and C2), keeps in the same plane and has no significant contributions to the corrugation. The nearest corrugated carbon atom was labeled as C3 in **Figure 5b**. The adsorption energy of Li$^+$ and K$^+$ of the corrugated graphene was calculated with equation of $E_{ads} = E_{tot} - E_R - E_{ion}$, where $E_{tot}$, $E_R$ and $E_{ion}$ are the total energy of ion/graphene, energy of corrugated graphene with different $R$ value, and energy of Li$^+$ or K$^+$ ion, respectively. From our calculations, the adsorption energy of graphene to electrolyte ions (*e.g.,* K$^+$, Li$^+$) decreased with increasing corrugations (**Figure 5c**). Since negative value of $E_{ads}$ represents adsorption and positive value for non-adsorption, this function relation make clearly that the interaction between ions (Li$^+$ or K$^+$) and graphene increased with increasing corrugation of graphene. With more corrugations, the distance between Li$^+$/K$^+$ and adsorption site of graphene was also decreased (**Figure 5d**). It should be noted that the difference between the interactions with K$^+$ and Li$^+$ (**Figure 5c**) is generated by their different ion radius (**Figure 5d**). The adsorption distance directly reflects the separation between electrolyte ions and the surface of electrode materials. According to *Eq.1*, the $C_{EDL}$ is therefore elevated upon the improvement on corrugation ratios. In principle, the ion concentration in Helmholtz layer should also be increased, which is similar to that found in graphene edge structures [19]. Besides, charge transfer between the C3 and adsorbed ions was slightly improved along with the corrugation (**Figure S11,** *e.g.,* from 0.891 to 0.895 of Li-C3), which will not be further discussed as the difference is just in such a small scale.

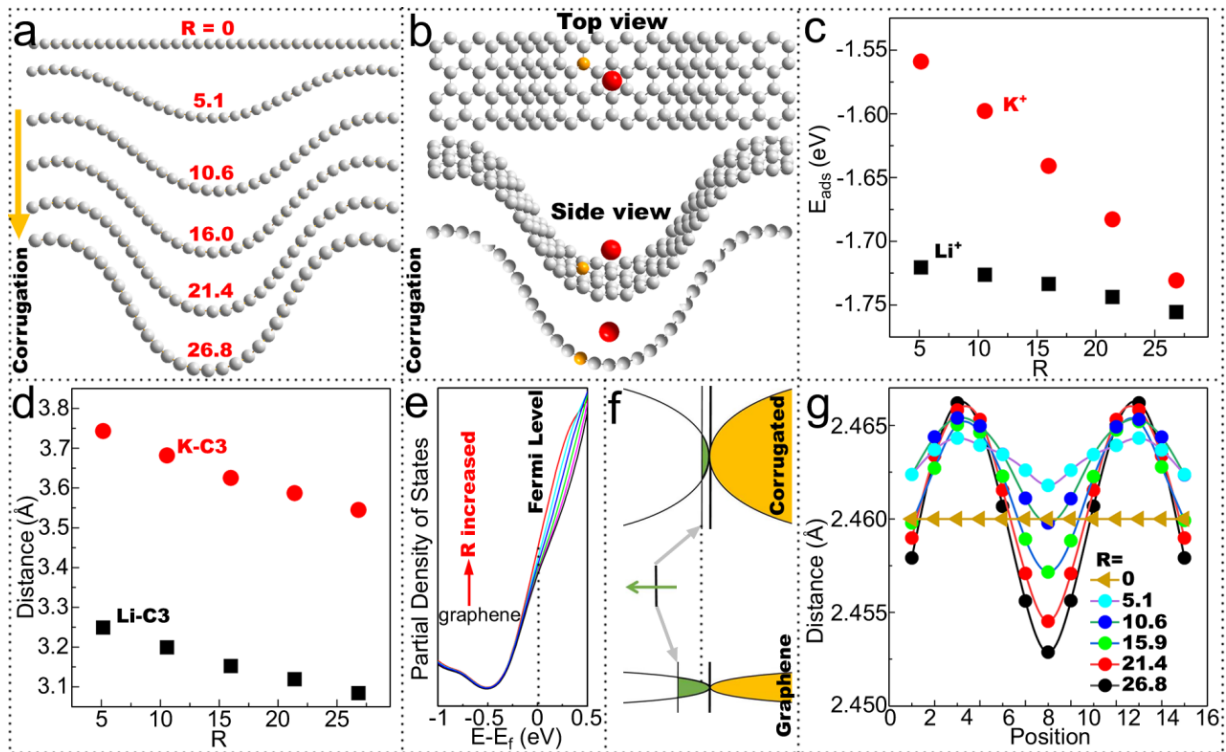

**Figure 5. Theoretical calculations. (a)** Side-view of optimized corrugated graphene. **(b)** top-view (upper) and side-view (lower) of ion/graphene, in which orange is labeled for the nearest corrugated C atom (C3) and red for adsorbed ion (Li$^+$ or K$^+$) on graphene surface. **(c)** Adsorption energy of Li$^+$ (black) and K$^+$ (red) on corrugated graphene surface. **(d)** Function relation between ion-C3 distance and corrugation ratio R. **(e)** PDOS of graphene and corrugated graphene with different R values (R = 0, 5.1, 10.6, 15.9, 21.4 and 26.8, increased from the bottom). **(f)** Diagram shows the electron transfer process from alkaline metals to graphene (bottom) and corrugated graphene (top). **(g)** Calculated distance between the next nearest neighboring carbon atoms along the wave direction (a structure example is showed in Figure S12).

The above promotion can be understood based on the improved DOS at the Fermi level. **Figure 5e** compares the partial density of states (PDOS) around the Fermi level of pure and corrugated graphene with different R ratios. Along with the improvement of corrugation R ratio, the DOS at the Fermi level increased accordingly. **Figure 5f** is a simplified diagram, showing how such improvement can be triggered by shifting electronic states to close to the Fermi level. When the charge transfers from the alkaline metal (or the adsorbed ions), the Dirac point of both graphene and corrugated graphene shifts to occupied states. Compared to that of pure graphene, the electron transferred from alkaline metals takes the lower energy state in corrugated graphene, corresponding to stronger adsorptions and giving better ion adsorptions discussed above. Besides, according to *Eq. 3*, the quantum capacitance of electrode materials is also improved because of the same reason. We further calculated the distance between the next nearest neighboring carbon atoms along with the wave direction (a side view structure example is shown in **Figure S12**). In this case, the width of the wave is steady, whilst the carbon distance along wave direction was significantly changed. Results of this calculation is shown in **Figure 5g**, which directly suggested that the average carbon distances in the crumples with R of 5.1, 10.6 and 15.9 were all larger than that of flat graphene (R = 0). Calculation of the point data suggested that all the sum of carbon distances in different crumples (R value from 5.1 to 26.8) were enlarged from that flat graphene (detailed data in **Figure S12**). In bump-like structures (the crumple is distorted with three-dimensional wave direction, **Figure 2d**), such enlargement should also works in different directions (along or perpendicular to the wave direction of one single crump), which directly improves the contacting area working for ion adsorptions. As a conclusion of the above DFT calculations, the enhanced adsorptions, decreased distance between ion and carbon surface, enhanced DOS at the Fermi level and the improved contacting area for ion adsorptions well explained why the energy storage capacity of graphene can be improved by corrugations.

**Conclusion**

In summary, we have prepared corrugated graphene by a rapid thermal expansion method. As-prepared graphene displays a BET surface area of ~340 $m^2 g^{-1}$. In energy storage applications, it delivered a high special capacitance of ~340 F g-1 at 5 mV $s^{-1}$ in aqueous three-electrode systems and ~343 F $g^{-1}$ at 1.7 A $g^{-1}$ in aqueous symmetric supercapacitor. In addition, a maximum energy density of 10.7 Wh $kg^{-1}$ at power density of 50 W $kg^{-1}$ was reached in aqueous electrochemical capacitor. Both the capacitance and power performance are comparable to and even better than that of high activated graphene materials with super-high BET surface area (*e.g.,* 3100 $m^2 g^{-1}$) and similar O-content. Our studies suggested that the considerable energy storage performance was attributed to the corrugation structures. These corrugations enhanced the DOS at the Fermi level of the graphene material, which shorten the distance to the adsorbed ions, elevated the binding ability, enhanced the quantum capacitance, and improved the contacting area for ion adsorptions. Alternative to the ultra-high BET surface area and rich porous structures, this study gives a different way to activate graphene materials for improved energy storage applications.

**SUPPLEMENTAL INFORMATION**
Supplemental Information can be found online at https://doi.org/xxxxxx.

**ACKNOWLEDGEMENT**
Financial supports from the National Natural Science Foundation of China (52172228) and UOW VC Fellowship are gratefully acknowledged. DFT calculations were supported by the Project of Hubei University of Arts and Science (XK2021024, 2020kypytd002).

## AUTHOTR CONTRIBUTION

**Longsheng Zhong**: conducted the main experiments under supervisions. **Chang Wu**: assisted experiments and characterizations under supervisions. **Xiaojing Zhu**: assisted experiments and characterizations under supervisions. **Shulai Lei**: performed DFT calculations and assisted supervisions. **Guijie Liang**: assisted DFT calculations and characterizations. **Sepidar Sayyar**: assisted characterizations. **Biao Gao**: assisted supervisions. **Liangxu Lin**: supervized the project, developed the fabrication techniques, conducted key preliminary experiments, analyzed data and drafted manuscript with assistance.

## DECLARATION OF INTERESTS

The authors declare no competing interests.

# Supplementary Information

## Fast Activation of Graphene with Corrugated Surface and its Role in Improved Aqueous Electrochemical Capacitors

*Longsheng Zhong,[1,#] Chang Wu,[2,3,#] Xiaojing Zhu,[4] Shulai Lei[5,*] Guijie Liang,[5] Sepidar Sayyar,[6,7] Biao Gao,[1,*] and Liangxu Lin[1,3,7,*]*

[1]Institute of Advanced Materials and Nanotechnology, The State Key Laboratory of Refractories and Metallurgy, Wuhan University of Science and Technology, Wuhan 430081, China
[2]Australia Institute for Innovative Materials, Innovation Campus, University of Wollongong, Squires Way, North Wollongong 2519, Australia
[3]The Straits Laboratory of Flexible Electronics (SLoFE), Fujian Normal University, Fuzhou 350117, China
[4]School of Mechanical and Electronic Engineering, China Jiliang University, Hangzhou 310018, China
[5]Hubei Key Laboratory of Low Dimensional Optoelectronic Materials and Devices, Hubei University of Arts and Science, Xiangyang 441053, China
[6]Australian National Fabrication Facility – Materials Node, Innovation Campus, University of Wollongong, Wollongong, NSW 2519, Australia
[7]ARC Centre of Excellence for Electromaterials Science, Intelligent Polymer Research Institute, Australia Institute for Innovative Materials, Innovation Campus, University of Wollongong, Squires Way, North Wollongong 2519, Australia

[#]Contribute equally to this work
*Correspondence: l.lin@wust.edu.cn (L. L.); sllei@hbuas.edu.cn (S. L.); gaobiao@wust.edu.cn (B. G.)

Figure S1-S12, Table S1, S2 and additional information

**Capacitance Calculation:** The specific capacitances (C, F g$^{-1}$) in three-electrode systems were calculated from CV curve with the equation:

$$C = \frac{\int I\, dV}{2 \cdot m \cdot U \cdot V}$$

where $\int I\, dV$, m, U and V are the integration area of CV curve, mass of active material, scan rate and potential window, respectively.

In two-electrode system, the specific capacitance (C, F g$^{-1}$,) was calculated from GCD with:

$$C = \frac{4I \cdot \Delta t}{m \cdot V}$$

where I, t, m and V are the constant discharge current, discharge time, total mass of active materials in both electrodes and potential window, respectively. Considering the pseudo-capacitance of the material, specific capacitance of the cell in aqueous media was also calculated from CV curves.

**Calculation of capacitive and diffusion-controlled fractions**: The calculation was based on the the well established approach with the following equation [20,30,31]:

$i(V) = k_1 v + k_2 v^{1/2}$

where the current response *i* at the potential V is contributed by both the capacitive effects (proportional to the scan rate *v*) and the diffusion-controlled process ($k_2 v^{1/2}$). $k_1$ and $k_2$ are the constant, which can be calculated from the a serious of CV plots and used to determine the fraction of the capacitive and diffusion-controlled process (see **Figure S5**). In our calculation, the scan rate of 5, 10, 20, 40 and 60 mV s$^{-1}$ was applied, while CV curves with higher scan rate were excluded.

**Additional information of Figure S1.** The a-G-375, annealed a-G-375 and control samples were very pure and only contained C and O elements (**Figures S1a,b,d,e**). Compared to the XPS of the raw GO material we reported previously (close to 50%) [27,28], the O was significantly reduced in both a-G-375 and annealed a-G-375-1000. The C1s XPS spectra of both a-G-375 and a-G-375-1000 are similar to that of highly reduced graphene materials [28], where the C-OH, C=O and O=C-OH are identified. XPS spectra of rGO-control-1 and rGO-control-2 are the same, with only difference on the C ratio. **Table S1** summarizes the ratio of different components in the C1s. In a-G-375, 18.56% C atoms were connected to O-containing group, which was reduced to 8.7% in annealed 3D-G-375-1000. Besides the sp2/sp3 carbon structures, a-G-375 also contains C-OH groups (blue shade), which was nearly disappeared in the annealed sample. The C ratio (mole) in a-G-375 and a-G-375-1000 was calculated as 84.34% and 90.4%, respectively; according to the C1s spectra (H atoms were not accounted). With identical method, the C ratio (mole) in rGO-control-1 and rGO-control-2 can also be calculated as 74.3 and 79.7, respectively (details in **Table S1**).

In **Figure S1c**, the G bands of both samples were found at 1585 and 1578 cm$^{-1}$, respectively. The red shift of the G band in annealed a-G-375-1000 means that the graphene has been reduced from a-G-375. It should be noted that the G band position in both a-G-375 and annealed a-G-375 are slightly high than we expected [28], which is reasonable when a very weak beam irradiation was applied to exclude the effect from thermal reductions [29]. In most reports, the thermal reduction induced by strong laser beam irradiation usually induced the huge red-shift of G band.

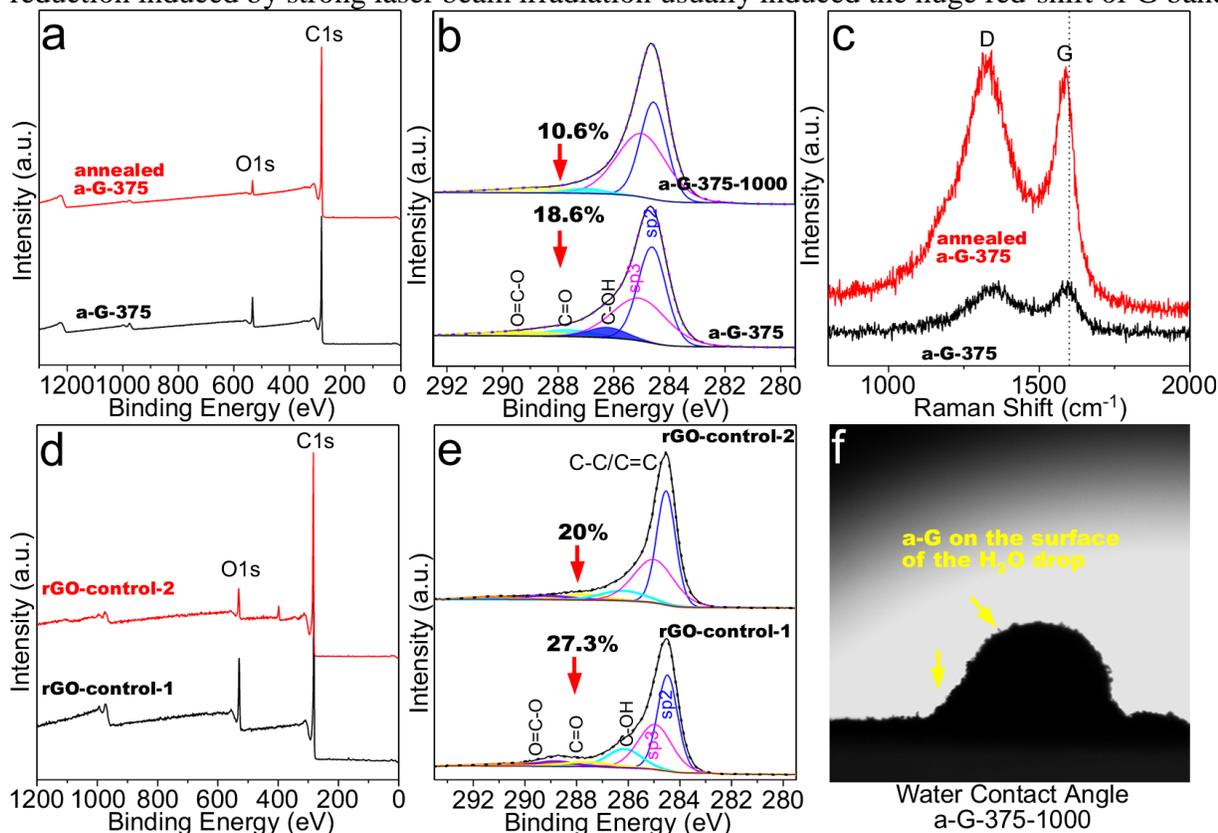

**Figure S1. (a-c)** XPS fully survey (a), C1s spectra (b) and Raman spectra (c) of a-G-375 and annealed a-G-375-1000. **(d-e)** XPS full survey (d) and C1s spectra (e) of rGO-control-1 and rGO-control-2. **(f)** Water contact angle of the annealed a-G-375-1000 shows some floating graphene particles on the surface of water drop. The annealed a-G-375-1000 should be highly hydrophobic.

**Table S1.** Contents of carbon and oxygen elements

| Materials | Mole Ratio (%) | | | | |
|---|---|---|---|---|---|
| | C-C/C=C | C-OH | C=O | -COOH | Carbon |
| a-G-375 | 81.44 | 6.36 | 5.93 | 6.27 | 84.34% |
| a-G-375-1000 | 91.3 | ~0 | 4.56 | 5.54 | 90.4% |
| r-GO-control-1 | 72.7 | 13.35 | 6.58 | 7.33 | 74.3% |
| r-GO-control-2 | 80 | 9.10 | 5.65 | 5.25 | 79.7% |

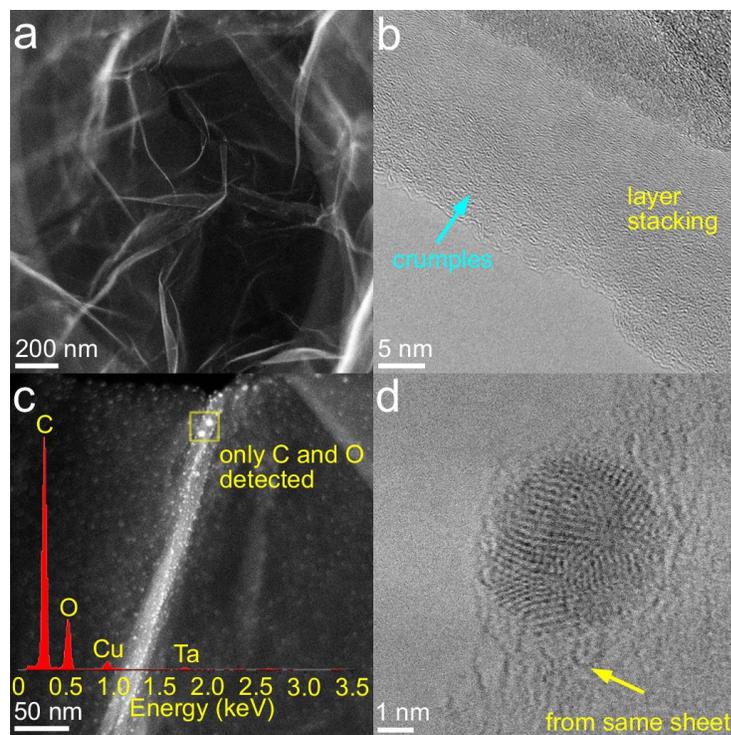

**Figure S2. (a-d)** TEM images of a-G-375 (a,c: dark field; b,d: bright field). The bright bump-like structures in (c) are in focus. Inset in (c) is the EDX spectrum of the selected dots which only gave C element (with some residual O). In the EDX spectrum, the Cu and Ta are from the holey carbon film grid and TEM holder, respectively. (d) The fine structure of one in-focused bump which is integrated with and from the rGO sheet.

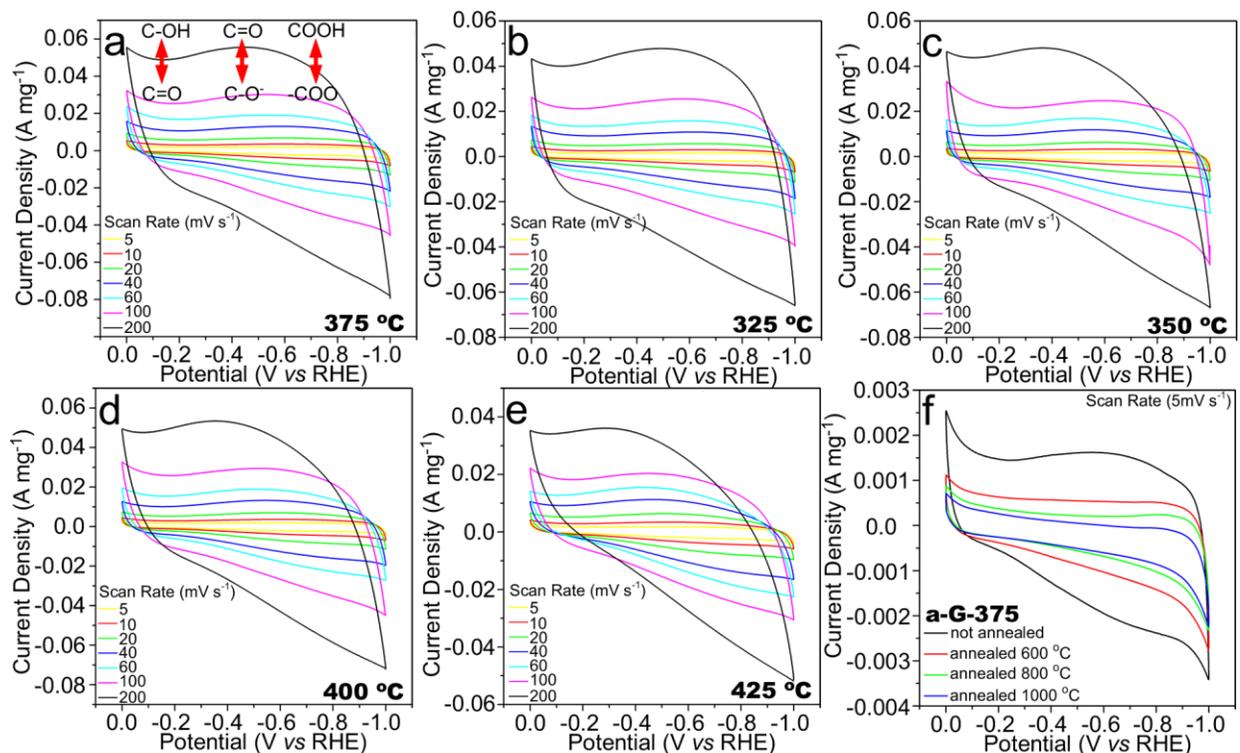

**Figure S3. CV curves recorded with three-electrode system in 6 M KOH.** (**a**) a-G-375. (**b**) a-G-325. (**c**) a-G-350. (**d**) a-G-400. (**e**) a-G-425. (**f**) a-G-375 and a-G-375-$x$ annealed at different temperatures.

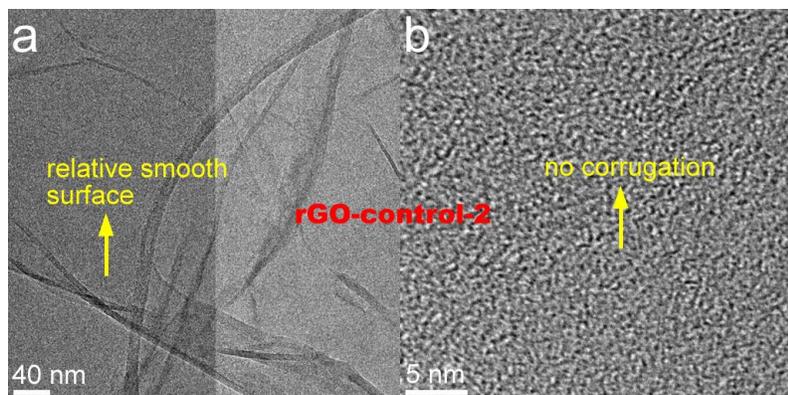

**Figure S4. (a-b)** Bright field TEM images of rGO-control-2. No clear surface corrugation was found on the rGO sheet, no matter the tilting of TEM holder was applied or not applied.

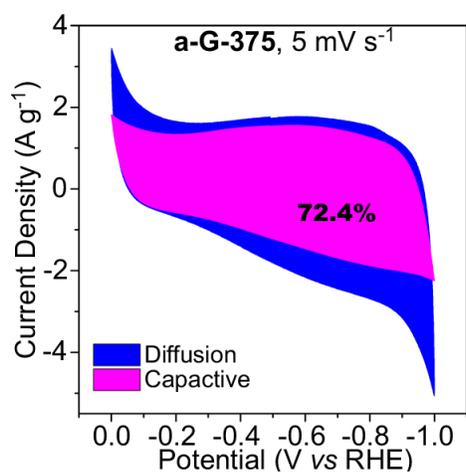

**Figure S5.** CV curve of a-G-375, with fractions of capacitive (magenta) and diffusion controlled contributor (blue). With identical method, the fractions of different samples at different scan rates were also calculated, and summarized in the bar char in **Figure 3g-i** and **Figure S6e-i**.

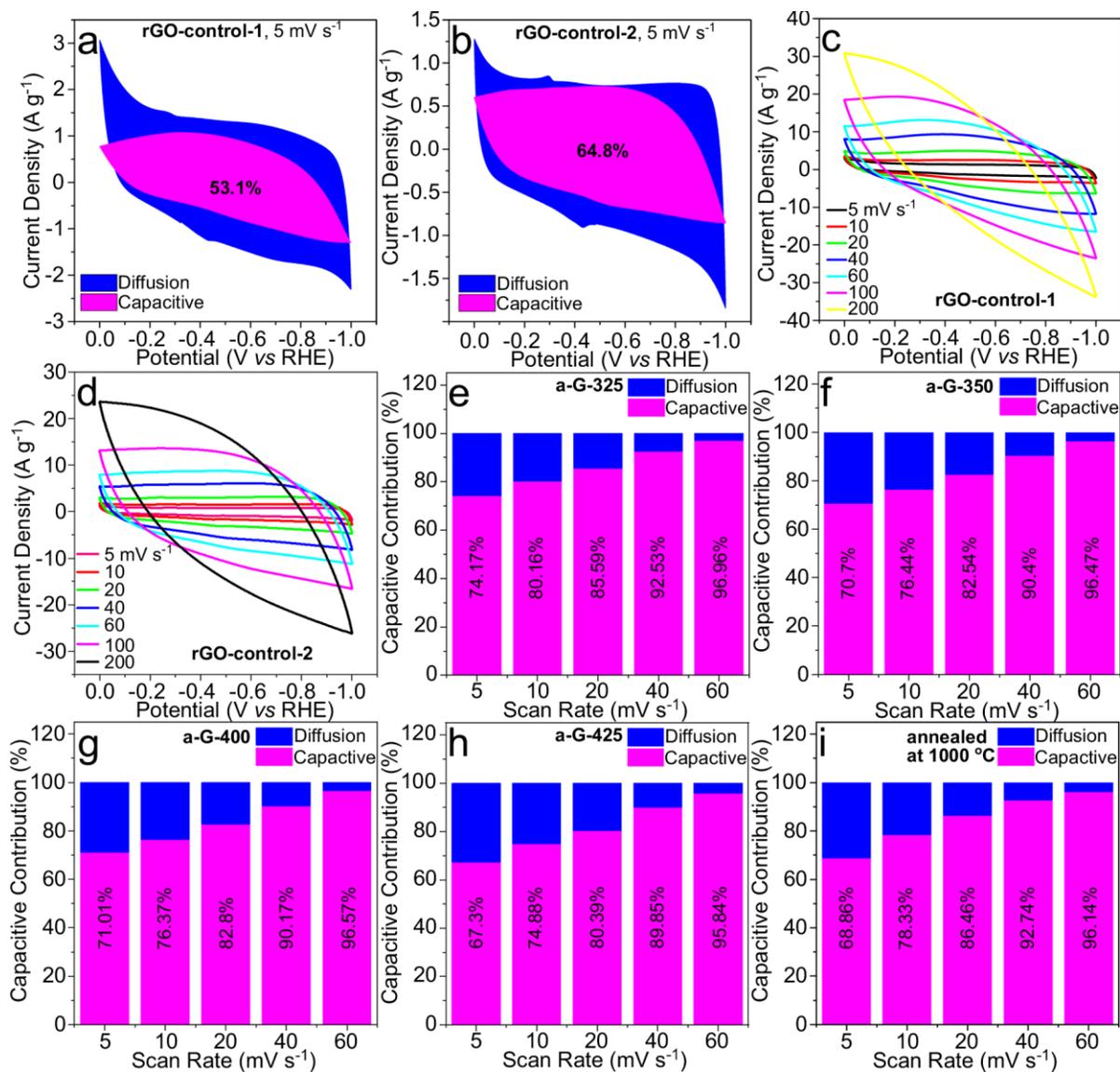

**Figure S6. (a,b)** CV curve of rGO-control-1 (a) and rGO-control-2 (b), with fractions of capacitive (magenta) and diffusion controlled contributor (blue). **(c,d)** CV curves of rGO-control-1 (c) and rGO-control-2 (d) at different scan rate. **(e-i)** The capacitive contribution at different scan rates of a-G-325 (e), a-G-350 (f), a-G-400 (g), a-G-425 (h) and a-G-350-1000 (i). Note: all the above measurements were performed in 6M KOH with three-electrode technique.

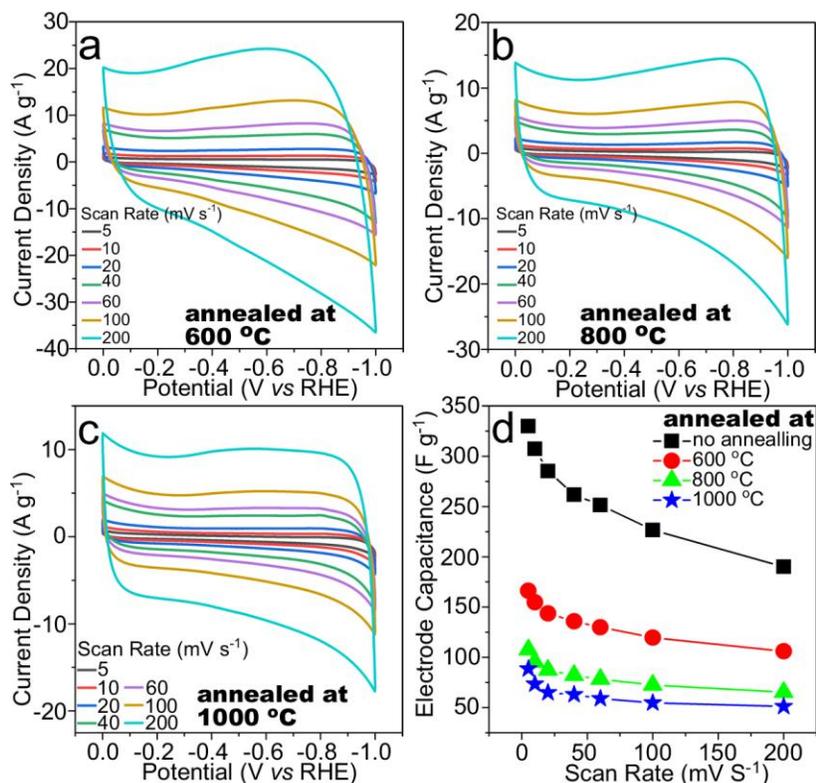

**Figure S7. (a-c)** CV curves of a-G-375 annealed at 600, 800 and 1000 °C for 2h, recorded with three-electrode system in 6 M KOH. (d) Plots of the electrode capacitance of different samples versus scan rate. The plot of a-G-375 is the same to that used in **Figure 3c**.

**Table S2.** Summary of special capacitance of different graphene materials in aqueous ECs. The power and energy densities were calculated according to the date reported in the literatures. $S_{BET}$ is the BET surface area.

| Materials | $S_{BET}$ (m² g⁻¹) | C-C/C=C (mole ratio) | Electrolyte (mol L⁻¹) | Current density (A g⁻¹) | Special capacitance (F g⁻¹) | Energy density (Wh kg⁻¹) / power density (W kg⁻¹) | Refs. |
|---|---|---|---|---|---|---|---|
| Few-layer graphene | - | 88.2 | KOH (6) | 0.5 | 113.2 | 2.52 / 200 | [s1] |
| Chemically modified graphene | 705 | - | KOH (5.5) | 0.01<br>0.02 | 135<br>128 | 4.69 / 5<br>4.44 / 10 | [s2] |
| Reduced graphene | 320 | 87.9 | KOH (30 wt.%) | 0.1 | 205 | - | [s3] |
| Reduced graphene | 326 | 88.1% | KOH (6) | 1 | 224.8 | 7.2 / 498.4 | [s4] |
| Graphene aerogels | 512 | 85.28% | KOH (6) | 0.05 | 128 | 4.44 / 25 | [s5] |
| Microwave exfoliated graphite | 463 | 73% | KOH (5) | 0.15<br>0.3<br>0.6 | 191<br>182<br>174 | 5.37 / 67.48<br>5.12 / 135.03<br>4.4 / 242.76 | [s6] |
| Crumpled graphene balls | 255 | 92.65% | KOH (5) | 0.1<br>2 | 150<br>118 | 3.33 / 40<br>2.62 / 800 | [s7] |
| Self-assembled graphene hydrogel | - | 83.8% | KOH (5) | 10 mV s⁻¹ (~1.75 A g⁻¹)<br>20 mV s⁻¹ (~3.04 A g⁻¹) | 175<br>152 | 6.08 / 218.9<br>5.28 / 437.76 | [s8] |
| Graphene oxide | 759 | 81.9% | KOH (6) | 0.5<br>2 | 154<br>15.7 | 5.35 / 250 | [s9] |
| Exfoliated graphene | 546 | 90.4% | | 2 | 42.8 | - | [s9] |
| Graphene nanoribbons | 378 | 98.1% | | 2 | 14.6 | - | [s9] |
| Reduced graphene aerogel | 38.7 | 85.75% | KOH (6) | 0.5 | 203.9 | 4.53 / 200 | [s10] |
| Chemically converted graphene liquid | 961.9 | - | $H_2SO_4$ (1) | 0.1 | 203.2 | 7.06 / 50 | [s11] |
| Laser-scribed | 1520 | - | $H_2SO_4$ (1) | 1 | 222 | 7.71 / 500 | [s12] |

| Materials | $S_{BET}$ (m² g⁻¹) | C-C/C=C (mole ratio) | Electrolyte (mol L⁻¹) | Current density (A g⁻¹) | Special capacitance (F g⁻¹) | Energy density (Wh kg⁻¹) / power density (W kg⁻¹) | Refs. |
|---|---|---|---|---|---|---|---|
| graphene | | | | | | | |
| N and S doped porous carbon nanosheets | 1744.58 | - | KOH (6) | 1 | 200 | 6.94/ 500 | [s13] |
| Pomegranate-like porous carbon | 1797 | 81.9% | H₂SO₄ (4) | 0.5 | 320 | 11.5 / 156 | [s14] |
| Porous 3D graphene sheets | 2308 | - | KOH (6) | 1 | 240 | - | [s15] |
| Polygonal nanomesh graphene | 1654 | 99.2% | KOH (6) | 1 | 245 | - | [s16] |
| Sponge-like porous carbons | 2580 | -- | H₂SO₄ (1) | 0.2 | 240 | 8.5 / 42 | [s17] |
| Activated sugar cane carbons | 1788 | 86.8% | H₂SO₄ (1) | 0.25 | 300 | 10 / 70 | [s18] |
| Hierarchically porous biochar | 2063 | - | Na₂SO₄ (1) | 1 | 291.6 | 10 / 500 | [s19] |
| 3D hollow porous graphene balls | 1871 | 82.6% | KOH (6) | 0.05<br>0.2 | 321<br>286 | - | [s20] |
| 3D highly porous graphene sheets | 1963 | 88.05% | KOH (6) | 1 | 172 | 8.7 / 600 | [s21] |
| Hierarchical porous graphene | 1810 | 96.09% | KOH (6) | 0.5 | 305 | 10.6 / 242.9 | [s22] |
| Activated reduced graphene oxide | 1720 | - | KOH (6) | 1 | 180 | 4.5 / 220 | [s23] |
| a-G-375 | ~340 | 84.34% | KOH (6) | 5 mV s⁻¹ (~1.7 A g⁻¹)<br>10 mV s⁻¹ (~3.0 A g⁻¹)<br>20 mV s⁻¹ (~4.9 A g⁻¹) | 342.7<br>297.7<br>247.1 | 11.9 / 214.2<br>10.3 / 372.1<br>8.6 / 617.8 | This work |

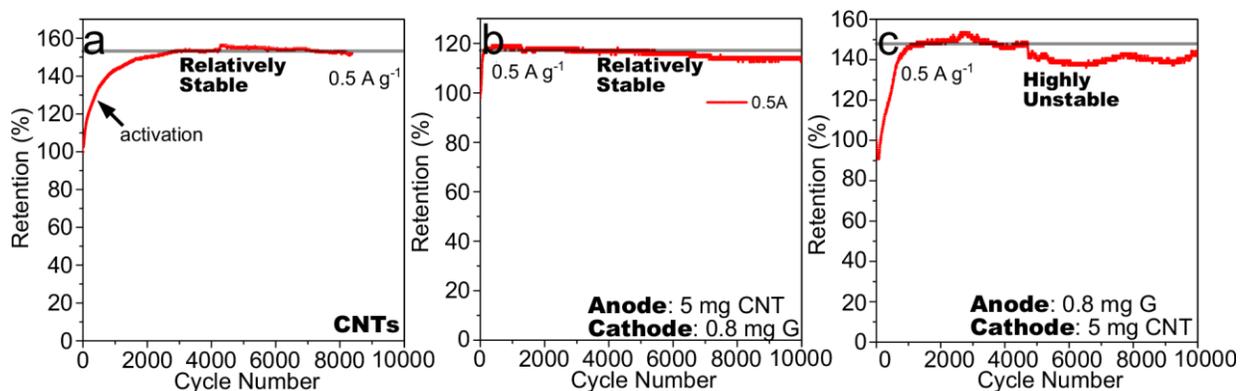

**Figure S8.** Cyclic test of symmetric and asymmetric ECs (6 M KOH electrolyte) at the current density of 0.5 A g⁻¹ with CP method. **(a)** Symmetric EC fabricated with pure CNTs. **(b)** Asymmetric EC with anode of 5 mg CNT and cathode of 0.8 mg activated graphene (G). **(c)** Asymmetric EC with anode of 0.8 mg activated graphene (G) and 5 mg CNT. CNTs were purchased from Hongwu New Material (Xuzhou) Co., Ltd.

**Additional information of Figure S8.** With identical method, we have also assembled different symmetric and asymmetric ECs. In the symmetric EC with active materials of CNTs, the capacitance of the device was relatively stable (**Figure S8a**). With 5 mg CNTs in anode and 0.8 mg activated graphene in cathode, the capacitance of the device was also relatively stable. However, when 0.8 mg activated graphene was used as the anode material, with cathode materials of 5 mg CNT, the measured capacitance was extremely unstable. Compared to commercial CNTs, the activated graphene materials here are more active, which turned to be relatively unstable in anode. Considering the surface structure of the activated graphene material, the corrosion in anode (also frequently observed in oxygen evolution reactions of carbon catalysts prepared in our group) and the existence of some O-containing groups may reduce the stability.

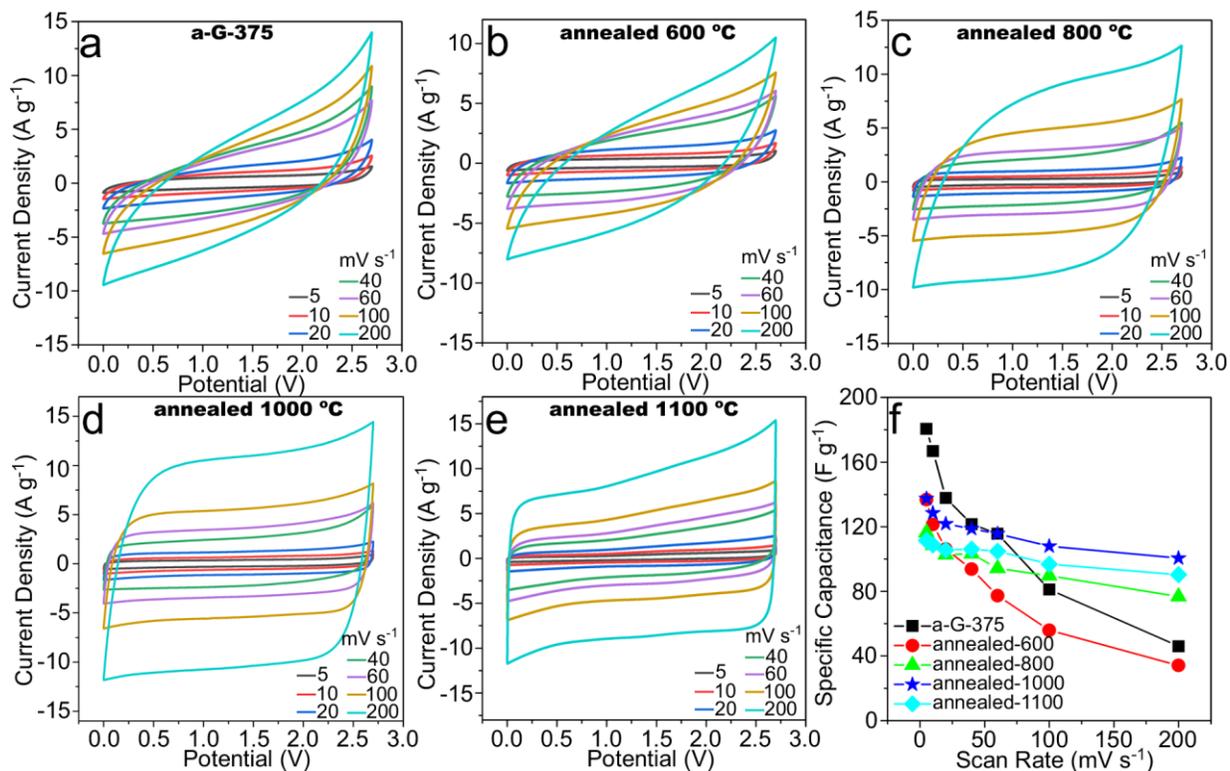

**Figure S9. (a-e)** CV curves of cell devices fabricated from different samples, including a-G-375 and samples annealed at 600, 800, 1000 and 1100 °C. **(f)** The comparison of the capacitance calculated from the CV curves.

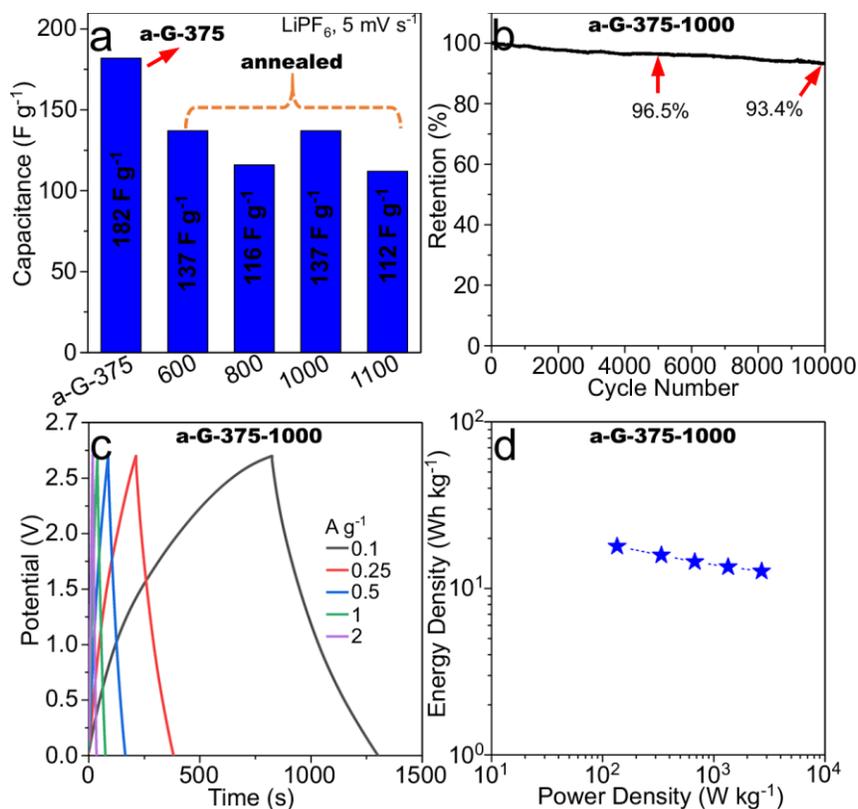

**Figure S10. Electrochemical performance of the cell assembled from a-G-375 and annealed a-G-375 with 1 M LiPF6 electrolyte**. **(a)** The comparison of specific capacitance of different samples at the scan rate of 5 mV s$^{-1}$. **(b)** Cyclic test at the current density of 0.5 A g$^{-1}$ with CP method of the cell assembled from a-G-375-1000. **(c)** GCD curves of the cell assembled from a-G-375-1000. **(d)** Ragone plot derived from figure c.

**Additional information of Figures S9-10**. **Figure S9a** shows the CV curves of the symmetric EC (in 1 M LiPF6, potential window of 2.7 V) fabricated from the annealed a-G-375-1000. Compared to other samples (a-G-375, a-G-375-600 and a-G-375-800, **Figure S9a-c**), the a-G-375-1000 (**Figure S9d**) and a-G-375-1100 (**Figure S9e**) have CV curves more close to rectangle shape, suggesting better capacitive behave and lower energy storage resistance. The plot of the specific capacitance (calculated from CV curves) against the scan rate also implied the better rate performance of a-G-375-1000 and a-G-375-1100 over that of others (**Figure S9e**). Compared to a-G-375-1100, the a-G-375-1000 showed higher specific capacitance (**Figure S9f, Figure S10a**), while their rate performances are nothing different. Specifically, the cell fabricated from a-G-375-1000 delivered specific capacitance of 137.4, 128.5, 121.8, 118.7, 115.6 and 107.9 F g$^{-1}$ at the scan rate of 5 (roughly equals to charge/discharge rate of 0.25 A g$^{-1}$), 10 (~0.48 A g$^{-1}$), 20 (~0.90 A g$^{-1}$), 40 (~1.76 A g$^{-1}$), 80 (~ 3.34 A g$^{-1}$) and 160 mV s$^{-1}$ (~6.4 A g$^{-1}$). In organic system, the cell performed similar cyclic stability to that in aqueous electrolyte. With GCD curves (cell assembled from a-G-375-1000, **Figure S10c**), we calculated the power density of 2.7 kW kg$^{-1}$ and energy density of 12.7 Wh kg$^{-1}$ at the rate of 2 A g$^{-1}$ (**Figure S10d**), which is not high enough comparing to the highly activated graphene with rich porous structures [9]. We have attempted various techniques (*e.g.,* to activate the cell at really low rate, to immerse the electrode piece into the electrolyte, to change to other organic electrolytes) to improve the power performance of the cell, but were not really successful.

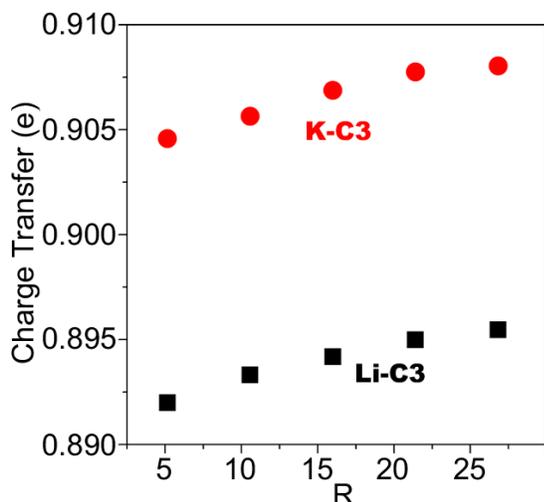

**Figure S11.** Function relation of the charge transfer number (between the C3 and adsorbed ion) and corrugation ratio R.

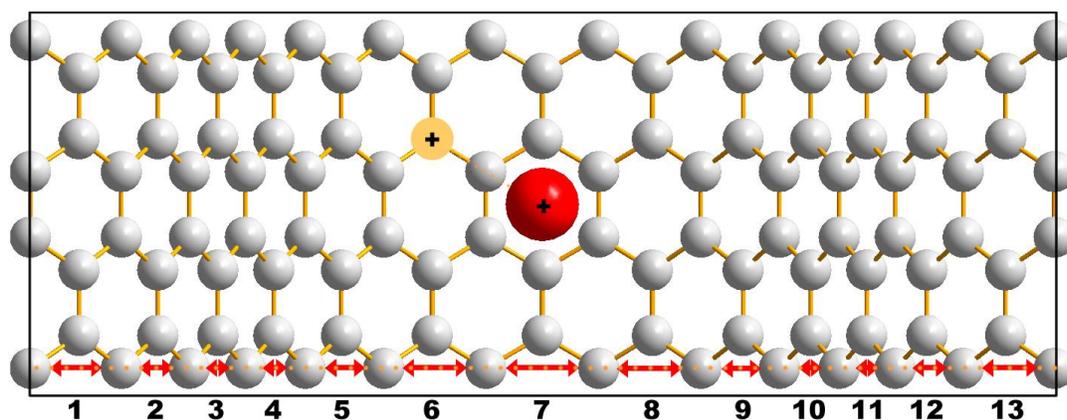

**Figure S12.** Side view structure shows the distance between C atom and the second-nearest neighbour in wave direction (R = 26.8). The Li$^+$/K$^+$ and C3 are indexed with red and yellow, respectively. The sum of carbon distances in the crumple with R values of 26.8, 21.4, 15.9, 10.6, 5.1 and 0 are calculated as 31.99038, 31.99871, 32.00426, 32.02291, 32.02249 and 31.98, respectively.

## Additional references